\documentclass[aps,prd,twocolumn,showpacs,amsmath,amssymb]{revtex4-1}
\usepackage{amsmath} \usepackage{graphicx} \usepackage{subfigure}
\usepackage{epstopdf} \usepackage{color} \usepackage{multirow}
\usepackage{setspace} \usepackage{overpic} \usepackage{amssymb}
\usepackage{ulem}
\usepackage[bookmarksnumbered, pdfstartview=FitH,colorlinks,urlcolor=blue, citecolor=blue,linkcolor=blue] {hyperref}
\usepackage{lineno}
\usepackage{bm}
\usepackage{rotating}
\usepackage{xcolor}%0824
\usepackage{makecell}
\usepackage{mathtext}
\usepackage{mathrsfs}
\usepackage{overpic}
\usepackage[T1]{fontenc}
\usepackage{lmodern}
\usepackage[utf8]{inputenc}
\let\oldequation\equation
\let\oldendequation\endequation
\renewenvironment{equation}
 {\linenomathNonumbers\oldequation}
 {\oldendequation\endlinenomath}

\begin{document}
%%===========================================
%%  For revised
%%===========================================
\definecolor{boslv}{rgb}{0.0, 0.65, 0.58}%persiangreen
\definecolor{Munsell}{HTML}{00A877}
\newcommand{\psip}{\psi^{'}}
\newcommand{\psipp}{\psi(3686)}

\newcommand{\Br}{\mathcal{B}}
\newcommand{\too}{\rightarrow}
\newcommand{\del}{\color{red}\sout}
\newcommand{\new}{\color{blue}\uwave}

%----------------------
\def\ifb{\mbox{fb$^{-1}$}}%  Inverse femtobarns.
\def\ipb{\mbox{pb$^{-1}$}}%  Inverse picobarns.
\def\inb{\mbox{nb$^{-1}$}}%  Inverse nanobarns.
\newcommand{\lamcplamcm}{\Lambda_{c}^{+}\bar{\Lambda}_{c}^{-}}
\newcommand{\lambdacp}{\Lambda_{c}^{+}}
\newcommand{\lambdacm}{\bar{\Lambda}_{c}^{-}}
%%************************************************************
%% Shared ones
%%************************************************************
%\newcommand{\BR}{\mathcal{B}}
\newcommand{\afs}{\alpha_s}
\newcommand{\bgp}{\beta\gamma}
\newcommand{\eff}{\varepsilon}
\newcommand{\sintht}{\sin{\theta}}
\newcommand{\costht}{\cos{\theta}}
\newcommand{\dedx}{dE/dx}

%%%%%%%%%%%%%%%%%%%%%%%%%%%%%%%%%%%%%%%%%%event selection
\newcommand{\probfc}{Prob_{\chi^2}}
\newcommand{\probpi}{Prob_{\pi}}
\newcommand{\probka}{Prob_{K}}
\newcommand{\probpr}{Prob_{p}}
\newcommand{\proball}{Prob_{all}}

%%%%%%%%%%%%%%%%%%%%%%%%%%%%%%%%%%%%%%%%%%charmonium
\newcommand{\chicJ}{\chi_{cJ}}
\newcommand{\gchicJ}{\gamma\chi_{cJ}}
\newcommand{\gchica}{\gamma\chi_{c0}}
\newcommand{\gchicb}{\gamma\chi_{c1}}
\newcommand{\gchicc}{\gamma\chi_{c2}}
\newcommand{\hc}{h_c(^1p_1)}
\newcommand{\qqb}{q\bar{q}}
\newcommand{\uub}{u\bar{u}}
\newcommand{\ddb}{d\bar{d}}
\newcommand{\SSB}{\Sigma^+\bar{\Sigma}^-}
\newcommand{\LLB}{\Lambda\bar{\Lambda}}
\newcommand{\ccb}{c\bar{c}}

%%************************************************************
%% Variables: decay modes of psiprime
%%************************************************************
\newcommand{\psipto}{\psi^{\prime}\rightarrow \pi^+\pi^- J/\psi}
\newcommand{\ptomm}{J/\psi\rightarrow \mu^+\mu^-}
\newcommand{\ppp}{\pi^+\pi^- \pi^0}
\newcommand{\pip}{\pi^+}
\newcommand{\pim}{\pi^-}
\newcommand{\kap}{K^+}
\newcommand{\kam}{K^-}
\newcommand{\ks}{K^0_s}
\newcommand{\pbar}{\bar{p}}
\newcommand{\jp}{J/\psi\rightarrow \gamma\pi^0}
\newcommand{\je}{J/\psi\rightarrow \gamma\eta}
\newcommand{\jep}{J/\psi\rightarrow \gamma\eta^{\prime}}

%%%%%%%%%%%%%%%%%%%%%%%%%%%%%%%%%%%%%%% 2prongs
\newcommand{\LL}{\ell^+\ell^-}
\newcommand{\ee}{e^+e^-}
\newcommand{\MM}{\mu^+\mu^-}
\newcommand{\GG}{\gamma\gamma}
\newcommand{\TT}{\tau^+\tau^-}
\newcommand{\pp}{\pi^+\pi^-}
\newcommand{\kk}{K^+K^-}
\newcommand{\ppb}{p\bar{p}}
\newcommand{\gpp}{\gamma \pi^+\pi^-}
\newcommand{\gkk}{\gamma K^+K^-}
\newcommand{\gppb}{\gamma p\bar{p}}
\newcommand{\ggee}{\gamma\gamma e^+e^-}
\newcommand{\gguu}{\gamma\gamma\mu^+\mu^-}
\newcommand{\ggll}{\gamma\gamma l^+l^-}
\newcommand{\ppee}{\pi^+\pi^- e^+e^-}
\newcommand{\ppuu}{\pi^+\pi^-\mu^+\mu^-}
\newcommand{\etap}{\eta^{\prime}}
\newcommand{\gpi}{\gamma\pi^0}
\newcommand{\geta}{\gamma\eta}
\newcommand{\getap}{\gamma\etap}
%%%%%%%%%%%%%%%%%%%%%%%%%%%%%%%%%%%%%%% 4prongs
\newcommand{\pppp}{\pi^+\pi^-\pi^+\pi^-}
\newcommand{\ppkk}{\pi^+\pi^-K^+K^-}
\newcommand{\pppr}{\pi^+\pi^-p\bar{p}}
\newcommand{\kkkk}{K^+K^-K^+K^-}
\newcommand{\kskp}{K^0_s K^+ \pi^- + c.c.}
\newcommand{\ppkp}{\pi^+\pi^-K^+ \pi^- + c.c.}
\newcommand{\ksks}{K^0_s K^0_s}
\newcommand{\dphi}{\phi\phi}
\newcommand{\phikk}{\phi K^+K^-}
\newcommand{\ppeta}{\pi^+\pi^-\eta}
\newcommand{\gpppp}{\gamma \pi^+\pi^-\pi^+\pi^-}
\newcommand{\gppkk}{\gamma \pi^+\pi^-K^+K^-}
\newcommand{\gpppr}{\gamma \pi^+\pi^-p\bar{p}}
\newcommand{\gkkkk}{\gamma K^+K^-K^+K^-}
\newcommand{\gkskp}{\gamma K^0_s K^+ \pi^- + c.c.}
\newcommand{\gppkp}{\gamma \pi^+\pi^-K^+ \pi^- + c.c.}
\newcommand{\gksks}{\gamma K^0_s K^0_s}
\newcommand{\gphiphi}{\gamma \phi\phi}

%%%%%%%%%%%%%%%%%%%%%%%%%%%%%%%%%%%%%%% 6prongs
\newcommand{\tpp}{3(\pi^+\pi^-)}
\newcommand{\tppkk}{2(\pi^+\pi^-)(K^+K^-)}
\newcommand{\pptkk}{(\pi^+\pi^-)2(K^+K^-)}
\newcommand{\tkk}{3(K^+K^-)}
\newcommand{\gtpp}{\gamma 3(\pi^+\pi^-)}
\newcommand{\gtppkk}{\gamma 2(\pi^+\pi^-)(K^+K^-)}
\newcommand{\gpptkk}{\gamma (\pi^+\pi^-)2(K^+K^-)}
\newcommand{\gtkk}{\gamma 3(K^+K^-)}

%%************************************************************
%%  Variables and Formula
%%************************************************************
\newcommand{\psp}{\psi(3686)}
\newcommand{\jpsi}{J/\psi}
\newcommand{\ar}{\rightarrow}
\newcommand{\lra}{\longrightarrow}
\newcommand{\jpsito}{J/\psi \rightarrow }
\newcommand{\ptoppjp}{J/\psi \rightarrow\pi^+\pi^- J/\psi}
\newcommand{\pspto}{\psi^\prime \rightarrow }
\newcommand{\ptop}{\psi'\rightarrow\pi^0 J/\psi}
\newcommand{\ptoeta}{\psi'\rightarrow\eta J/\psi}
\newcommand{\ecto}{\eta_c \rightarrow }
\newcommand{\ecpto}{\eta_c^\prime \rightarrow }
\newcommand{\xto}{X(3594) \rightarrow }
\newcommand{\chicJto}{\chi_{cJ} \rightarrow }
\newcommand{\chiczto}{\chi_{c0} \rightarrow }
\newcommand{\chicoto}{\chi_{c1} \rightarrow }
\newcommand{\chictto}{\chi_{c2} \rightarrow }
\newcommand{\pspp}{\psi^{\prime\prime}}
\newcommand{\ptochic}{\psi(2S)\ar \gamma\chi_{c1,2}}
\newcommand{\ppjpsi}{\pi^0\pi^0 J/\psi}
\newcommand{\utoeta}{\Upsilon^{\prime}\ar\eta\Upsilon}
\newcommand{\ww}{\omega\omega}
\newcommand{\wf}{\omega\phi}
\newcommand{\ff}{\phi\phi}
\newcommand{\npsp}{N_{\psp}}
\newcommand{\llb}{\Lambda\bar{\Lambda}}
\newcommand{\llbpi}{\llb\pi^0}
\newcommand{\llbeta}{\llb\eta}
\newcommand{\ppi}{p\pi^-}
\newcommand{\pbpi}{\bar{p}\pi^+}
\newcommand{\lamb}{\bar{\Lambda}}
%%=================================================
%% abbreviated commands of LaTeX
%%=================================================
\def\ctup#1{$^{\cite{#1}}$}
\newcommand{\bfg}{\begin{figure}}
\newcommand{\efg}{\end{figure}}
\newcommand{\bitm}{\begin{itemize}}
\newcommand{\eitm}{\end{itemize}}
\newcommand{\bnum}{\begin{enumerate}}
\newcommand{\enum}{\end{enumerate}}
\newcommand{\btbl}{\begin{table}}
\newcommand{\etbl}{\end{table}}
\newcommand{\btbu}{\begin{tabular}}
\newcommand{\etbu}{\end{tabular}}
\newcommand{\bcl}{\begin{center}}
\newcommand{\ecl}{\end{center}}
\newcommand{\bbt}{\bibitem}
\newcommand{\beq}{\begin{equation}}
\newcommand{\eeq}{\end{equation}}
\newcommand{\beqr}{\begin{eqnarray}}
\newcommand{\eeqr}{\end{eqnarray}}
%%===========================================
%%  color setting
%%===========================================
\newcommand{\red}{\color{red}}
\newcommand{\blue}{\color{blue}}
\newcommand{\yellow}{\color{yellow}}
\newcommand{\green}{\color{green}}
\newcommand{\purple}{\color{purple}}
\newcommand{\brown}{\color{brown}}
\newcommand{\black}{\color{black}}

\title{\boldmath Search for a bound state of $\Lambda_{c}\bar{\Sigma}_{c}$ near threshold }
%% Saved at => 2025-05-29
%\author{Author list}
\author{%%
\begin{small}
\begin{center}
M.~Ablikim$^{1}$, M.~N.~Achasov$^{4,c}$, P.~Adlarson$^{77}$, X.~C.~Ai$^{82}$, R.~Aliberti$^{36}$, A.~Amoroso$^{76A,76C}$, Q.~An$^{73,59,a}$, Y.~Bai$^{58}$, O.~Bakina$^{37}$, Y.~Ban$^{47,h}$
% ...
\\
\vspace{0.2cm}
(BESIII Collaboration)\\
\vspace{0.2cm} {\it
$^{1}$ Institute of High Energy Physics, Beijing 100049, People's Republic of China\\
% ...
$^{4}$ Budker Institute of Nuclear Physics SB RAS (BINP), Novosibirsk 630090, Russia\\
% ...
$^{36}$ Johannes Gutenberg University of Mainz, Johann-Joachim-Becher-Weg 45, D-55099 Mainz, Germany\\
$^{37}$ Joint Institute for Nuclear Research, 141980 Dubna, Moscow region, Russia\\
% ...
$^{47}$ Peking University, Beijing 100871, People's Republic of China\\
% ...
$^{58}$ Southeast University, Nanjing 211100, People's Republic of China\\
$^{59}$ State Key Laboratory of Particle Detection and Electronics, Beijing 100049, Hefei 230026, People's Republic of China\\
% ...
$^{73}$ University of Science and Technology of China, Hefei 230026, People's Republic of China\\
% ...
$^{76}$ University of Turin and INFN, (A)University of Turin, I-10125, Turin, Italy; (B)University of Eastern Piedmont, I-15121, Alessandria, Italy; (C)INFN, I-10125, Turin, Italy\\
$^{77}$ Uppsala University, Box 516, SE-75120 Uppsala, Sweden\\
% ...
$^{82}$ Zhengzhou University, Zhengzhou 450001, People's Republic of China\\
\vspace{0.2cm}
$^{a}$ Deceased\\
% ...
$^{c}$ Also at the Novosibirsk State University, Novosibirsk, 630090, Russia\\
% ...
$^{h}$ Also at State Key Laboratory of Nuclear Physics and Technology, Peking University, Beijing 100871, People's Republic of China\\
% ...
}\end{center}
\vspace{0.4cm}
\end{small}
}
%% ends here %%

\date{\today}
% \linenumbers

\begin{abstract}
 We  search for a possible $\Lambda_{c} \bar{{\Sigma}}_{c}$ bound state, denoted as $H_{c}^{\pm}$, via the $ \ee \to \pi^{+} \pi^{-} \lamcplamcm$ process for the first time. 
 This analysis utilizes 207.8 and 159.3~$\ipb$ of $\ee$  annihilation data at the center-of-mass energies of 4918.02 and 4950.93~MeV, respectively, collected with the BESIII detector at the BEPCII collider.
 No statistically significant signal is observed. 
 The upper limits of the product of Born cross section and branching fraction $\sigma(\ee \to \pi^{+}  H_c^{-} + c.c.) \times \mathcal{B}(H_c^{-} \rightarrow   \pi^{-}\lamcplamcm)$ 
 at a 90\% confidence level are reported at each energy point and for various $H_{c}$ mass hypotheses (4715, 4720, 4725, 4730, and 4735~MeV/$c^{2}$) and  widths (5, 10, or 20~MeV), with the upper limits ranging from 1.1 pb to 6.4 pb.

\end{abstract}

\maketitle
	%%%%%%%%%%%%%%%%%%%%%%%%%%%%%%%%%%%%%%%%
	%          1. Introduction
	%%%%%%%%%%%%%%%%%%%%%%%%%%%%%%%%%%%%%%%%
\section{Introduction}

Over the past  decades, numerous hadron states and resonant structures have emerged in experimental data. 
Many of these discoveries deviate from the conventional quark-antiquark or three-quark hadron expectations   from the quark model, and are likely evidence for exotic states such as multiquark states, hadronic molecules, hybrids, or glueballs~\cite{xyz-ycz}.
Understanding their nature is crucial for advancing our knowledge of Quantum Chromodynamics (QCD). 
A notable feature of these states is their tendency to cluster near the thresholds of heavy-antiheavy hadron pairs.
Prominent examples include the $X(3872)$ (also known as $\chi_{c1}(3872)$)~\cite{xyz1} and the $Z_{c}(3900)^{\pm}$ (also known as $T_{c\bar{c}1}(3900)^{\pm}$)~\cite{xyz2}, which are situated near the $D\bar{D}^{\ast}$ threshold. These resonant structures have been extensively studied in various models, with many being interpreted as molecular states within their respective systems~\cite{molecular}.
Among the related structures predicted to exist near the thresholds of heavy-antiheavy hadron pairs is the $ \Lambda_c \bar{\Sigma}_c$ bound state a particularly intriguing candidate, consisting of a charmed baryon and an anti-charmed baryon.
It is predicted to exist with   $(I,S) = (1,0)$ and $J^{PC} = (0,1)^{-\pm}$~\cite{dxk}. However, no such bound state has been experimentally observed to date.

In this paper,  the process $ \ee \to \pi^{+} \pi^{-} \lamcplamcm$ has been used to search for a bound state of $\Lambda_{c}\bar{\Sigma}_{c}$, denoted $H_{c}^{\pm}$, decaying into $\pi^{\pm}\lamcplamcm$. Charge-conjugated modes are implied throughout this paper. The analysis utilizes two data samples at the center-of-mass (c.m.) energies $\sqrt{s}=$ 4918.02 and 4950.93~MeV, corresponding to integrated luminosities of $207.8 \pm 1.1$ and $159.3 \pm 0.9~\ipb$~\cite{cms}, respectively.

	%%%%%%%%%%%%%%%%%%%%%%%%%%%%%%%%%%%%%%%%%%%%%%%%%%%%%%%%%%%%%%%
	%          2. BESIII DETECTOR AND MONTE CARLO SIMULATION
	%%%%%%%%%%%%%%%%%%%%%%%%%%%%%%%%%%%%%%%%%%%%%%%%%%%%%%%%%%%%%%%
\section{BESIII DETECTOR AND MONTE CARLO SIMULATION}\label{sec2}
The BESIII detector~\cite{Ablikim:2009aa} records symmetric $\ee$ collisions provided by the BEPCII storage ring~\cite{CXYu_bes3}, which operates with a peak luminosity of $1.1 \times 10^{33} \text{cm}^{-2}\text{s}^{-1}$ in the c.m.~energy range from 1.84 to 4.95~GeV. The cylindrical core of the BESIII detector covers 93\% of the full solid angle and consists of a helium-based multilayer drift chamber (MDC), a plastic scintillator time-of-flight system (TOF), and a CsI(Tl) electromagnetic calorimeter (EMC), which are all enclosed in a superconducting solenoidal magnet providing a 1.0~T magnetic field. The solenoid is supported by an octagonal flux-return yoke with resistive plate counter muon identification modules interleaved with steel. The charged-particle momentum resolution at 1~GeV/$c$ is 0.5\%, and the $dE/dx$ resolution is 6\% for electrons from Bhabha scattering. The EMC measures photon energies with a resolution of 2.5\% (5\%) at 1~GeV in the barrel (end cap) region. The time resolution in the TOF barrel region is 68 ps, while that in the end cap region is 110~ps. The end cap TOF system was upgraded in 2015 using multi-gap resistive plate chamber technology, providing a time resolution of 60~ps~\cite{tof_a,tof_b,tof_c}.

Simulated data samples are produced with a {\sc geant4}-based~\cite{geant4} Monte Carlo (MC) package, which includes the geometric description of the BESIII detector and the detector response~\cite{detvis}. They are used to optimize the event selection criteria and estimate the signal efficiency and the level of background. The simulation models the beam-energy spread and initial-state radiation (ISR) in the $e^+e^-$ annihilation using the generator {\sc kkmc}~\cite{kkmc_a}. 
The inclusive MC sample, which consists of $\lamcplamcm$ events, $D_{(s)}$ production, $\psi$ states produced in initial state radiation processes, and continuum processes  $\ee \to q\bar{q} (q = u, d, s)$, are generated to estimate the potential background. Here, all the known decay modes of charmed hadrons and charmonia are modeled with {\sc evtgen}~\cite{evtg1,evtg2} using branching fractions (BFs) taken from the Particle Data Group  (PDG)~\cite{pdg2024}, while the remaining unknown decays are modeled with    {\sc lundcharm}~\cite{lundcharm_a,lundcharm_b}. Final state radiation from charged final state particles is incorporated using the  {\sc photos} package~\cite{photos}.
Given that the  mass thresholds for the $\pi\Lambda_{c}\bar{\Lambda}_{c}$  and $\Lambda_{c}\bar{\Sigma}_{c}$  systems are 4710~MeV/$c^{2}$  and  4740~MeV/$c^{2}$, respectively, we set the $H_{c}$ mass to the values 4715, 4720, 4725, 4730, and 4735~MeV/$c^{2}$, each with widths of 5, 10, or 20~MeV. For each of these 15 mass-width combinations, 60,000 MC events are simulated at each of the two energy points.
The signal MC samples  simulate the process $\ee \to \pi^{+} H_{c}^{-}$, $H_{c}^{-} \to \pi^{-}\lamcplamcm$, where the $\lambdacm$ decays generically while the $\lambdacp$ decays into  $pK^{-}\pi^{+}$.
The decays of the signal process are modeled with a uniform phase space distribution, except for $\lambdacp \to pK^{-}\pi^{+}$, which uses the result from a partial wave analysis~\cite{pkpi}.

	%%%%%%%%%%%%%%%%%%%%%%%%%%%%%%%%%%%%%%%%
	%          3. EVENT SELECTION
	%%%%%%%%%%%%%%%%%%%%%%%%%%%%%%%%%%%%%%%%
\section{DATA ANALYSIS}\label{sec3}

A partial reconstruction method is applied for the signal process $\ee \to \pi^{+} H_{c}^{-}$, $H_{c}^{-} \to \pi^{-}\lamcplamcm$ to achieve  a higher efficiency. 
The $\Lambda_{c}^{+}$ is reconstructed using the ``Golden Mode'' $\Lambda_{c}^{+} \to pK^{-}\pi^{+}$~\cite{golden}.
If an additional  $\pi$ is identified, the recoil mass of the $\pi$ is calculated; otherwise the kinematic information of the reconstructed $\Lambda_{c}^{+}$ will be used to identify the signal. 

Charged tracks detected in the  MDC are required to be within a polar angle ($\theta$) range of |cos$\theta$| < 0.93, where $\theta$ is defined with respect to the $z$-axis (the symmetry axis of the MDC). The distance of the closest approach to the interaction point (IP) for charged tracks is required to be $\pm10$~cm along the $z$-axis and 1~cm in the plane perpendicular to the beam. 
Particle identification (PID) is implemented by combining measurements of the specific  ionization energy loss in the MDC and the TOF between the IP and the  dedicated TOF detector system. Each charged track is assigned to be a  pion,  kaon, or proton, according to which assignment has the highest probability. 

The $\Lambda_{c}^{+}$ candidates are reconstructed by looping over all combinations of $pK^{-}\pi^{+}$  that satisfy a vertex fit, with a $\chi^{2}$ value from the vertex fit less than 200. 
We require the $pK^{-}\pi^{+}$ invariant mass, denoted as $M(pK^{-}\pi^{+})$, to be within the mass window [2.270, 2.300] GeV/$c^{2}$~\cite{fjh}.
If  multiple  $\Lambda_{c}^{+}$ candidates are found within a single event,  the one with the minimum $\chi^{2}$ value is retained.
For any remaining  $\pi^{+}$  or $\pi^{-}$ particles, a vertex fit is performed to the  $\Lambda_{c}^{+}\pi^{+}$ or $\Lambda_{c}^{+}\pi^{-}$  combinations, with a requirement that the $\chi^{2}$ of the vertex fit is also less than 200.
If multiple $\Lambda_{c}^{+}\pi^{+}$ or $\Lambda_{c}^{+}\pi^{-}$ combinations exist, we select the one with the minimum $\chi^{2}$.
If an event contains both $\Lambda_{c}^{+} \pi^+$ and $\Lambda_{c}^{+} \pi^-$ combinations, both are retained, as the signal process involves both $\pi^{+}H_{c}^{-}$ and $\pi^{-}H_{c}^{+}$ by charge conjugation. 
We select the combination with the best vertex fit to reduce the long-lived particles,  $\Lambda$ and $K_{S}$, in the $\Lambda_{c}$ decay channels.
Events that contain only $\Lambda_{c}^{+}$ and no other $\pi$ are labeled as ``$\Lambda_{c}$ Tag'', while those that include an extra $\pi$ are designated as ``$\Lambda_{c}\pi$ Tag''.
The distributions of  $M(pK^{-}\pi^{+})$ for the two tags are shown in Fig.~\ref{fig1}, where evident $\lambdacp$ signals are observed at both c.m. energies.

\begin{figure}[htbp]
\begin{center}
\begin{minipage}[t]{1.0\linewidth}
\includegraphics[width=0.8\textwidth]{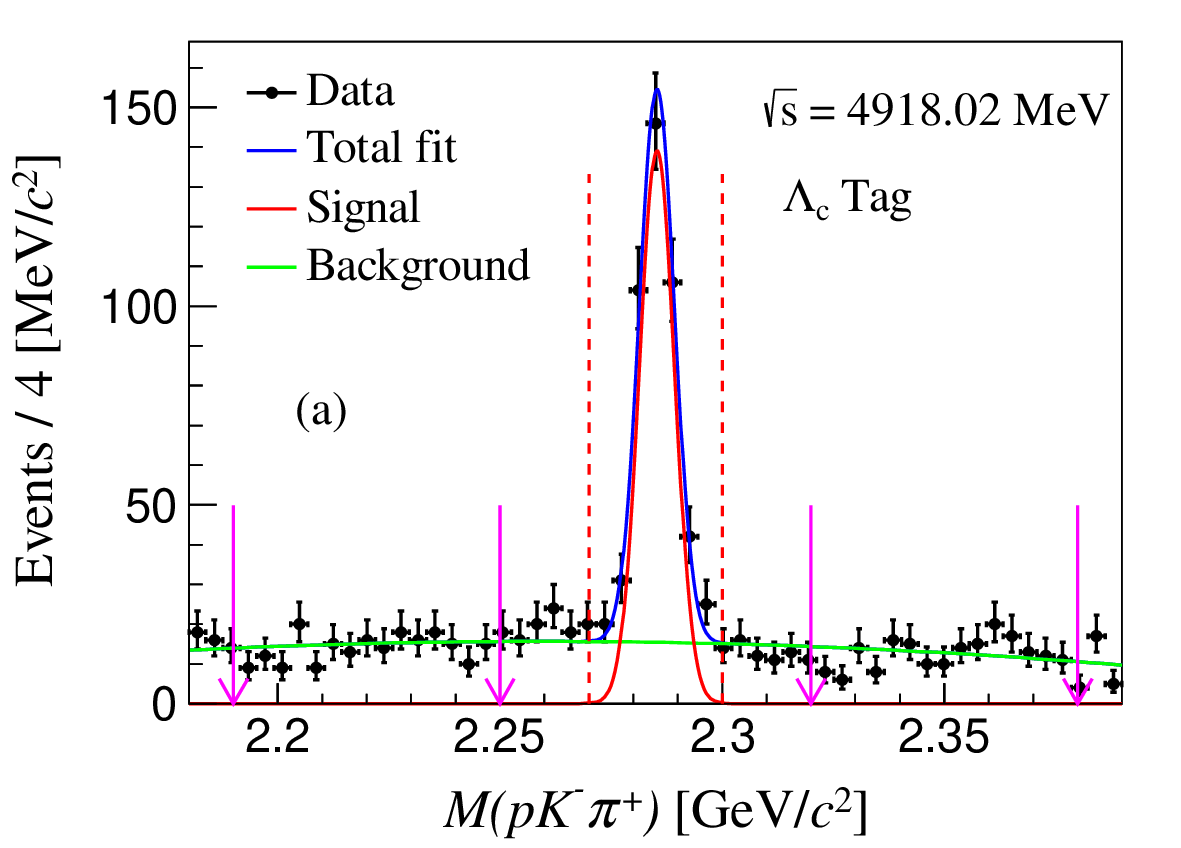}
\includegraphics[width=0.8\textwidth]{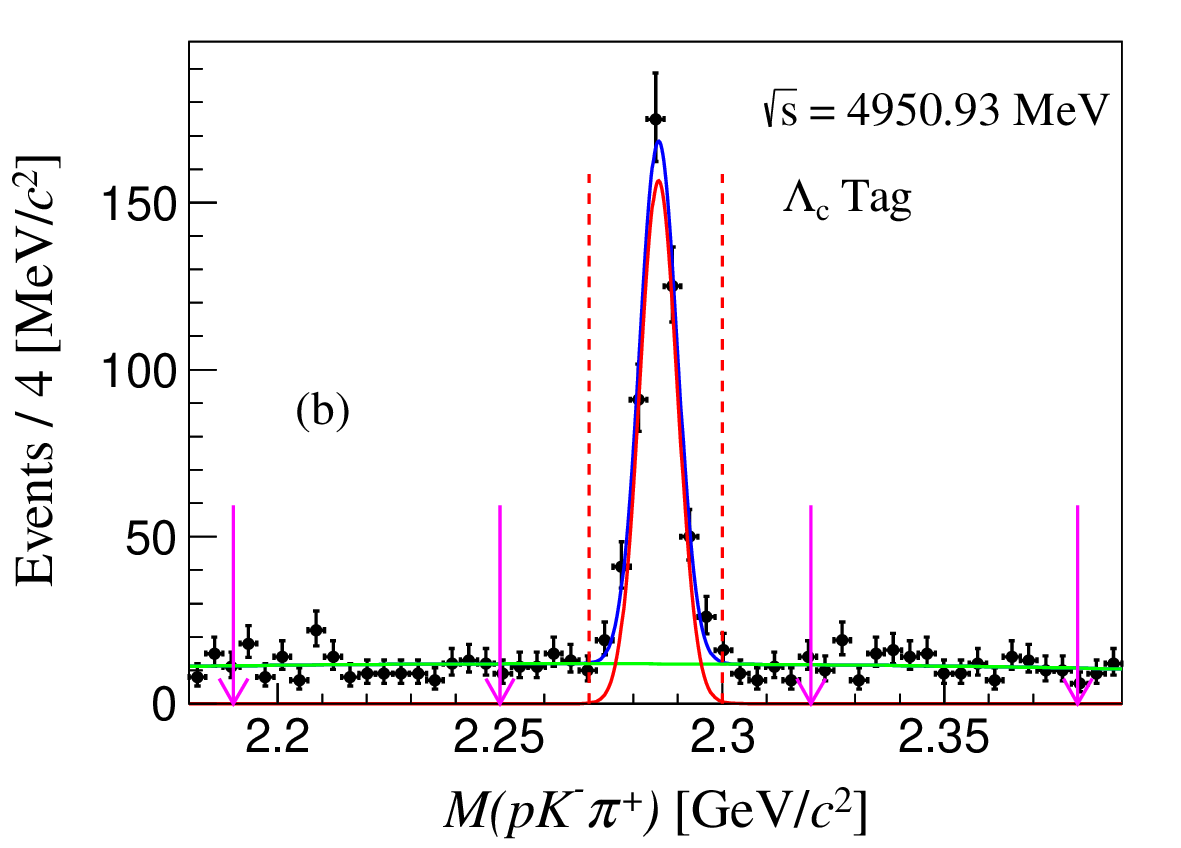}
\includegraphics[width=0.8\textwidth]{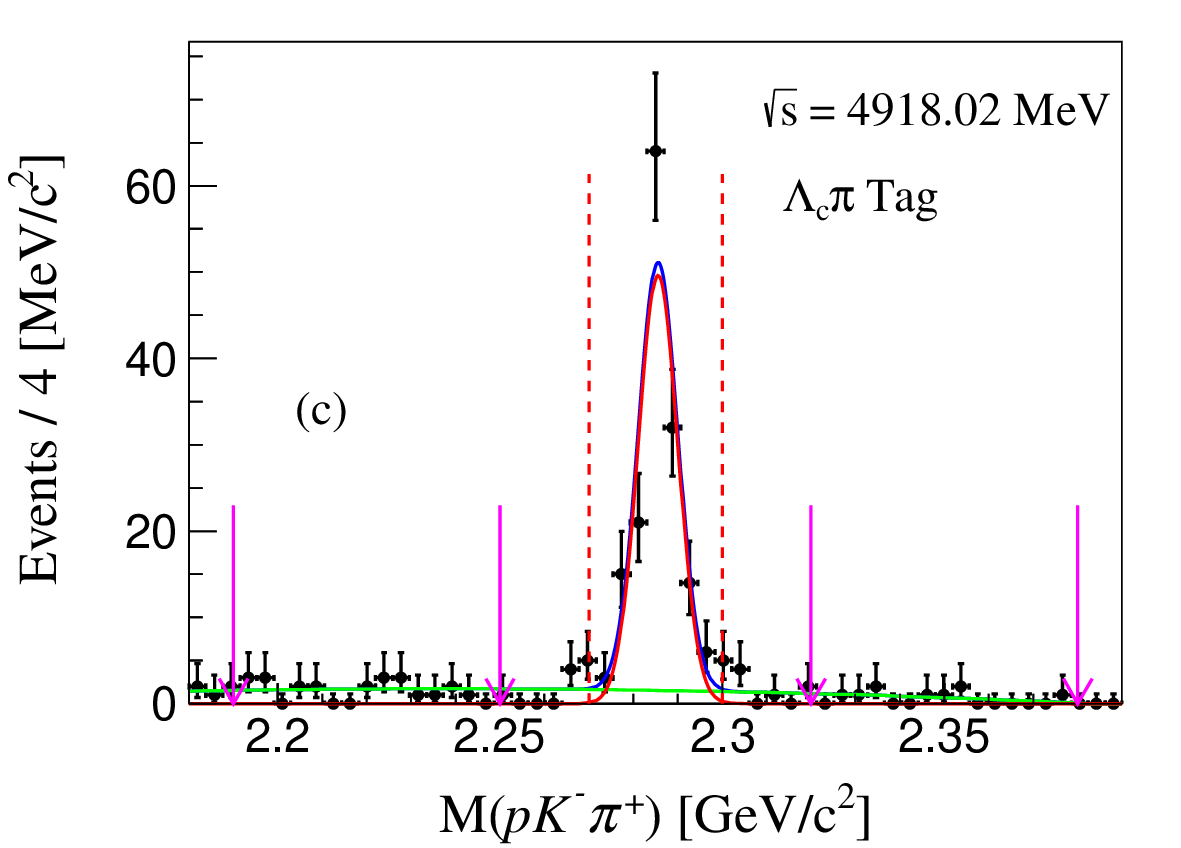}
\includegraphics[width=0.8\textwidth]{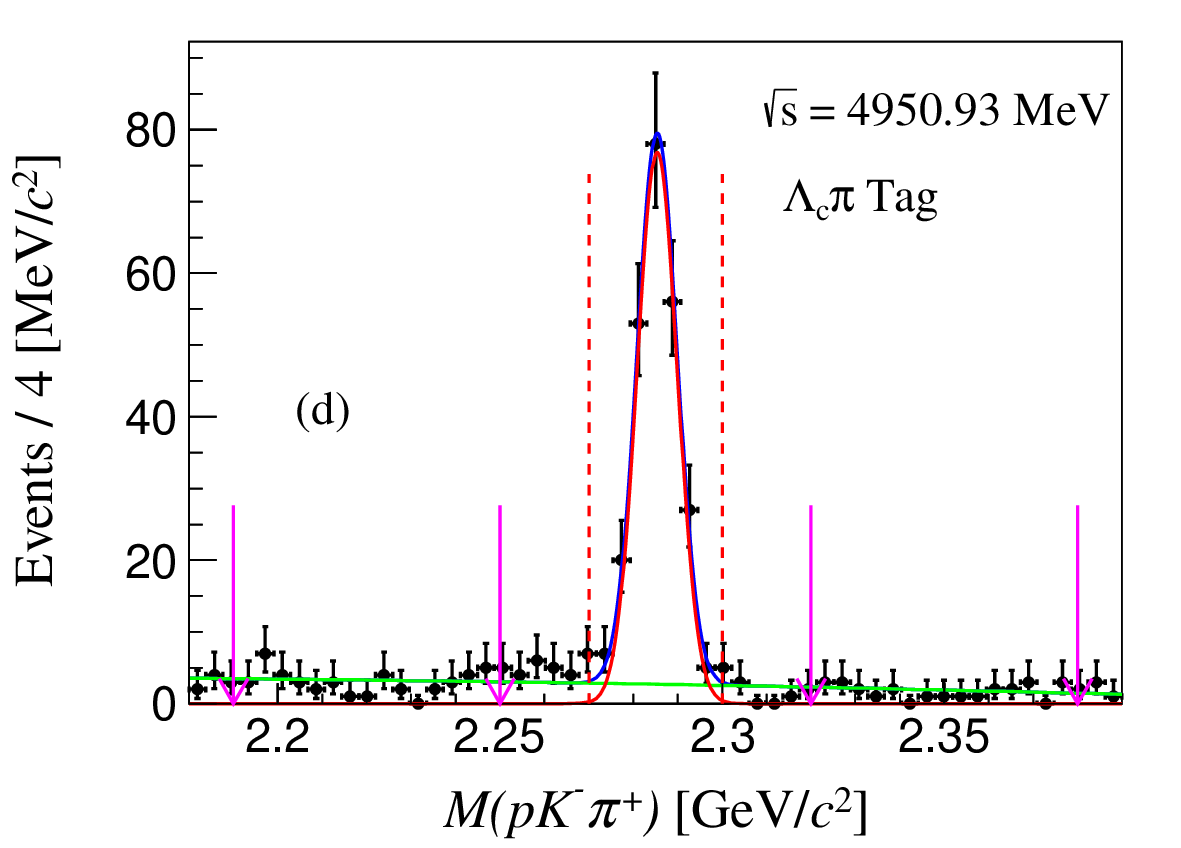}
\end{minipage}
\caption{The $M(pK^{-}\pi^{+})$ distributions for ``$\Lambda_{c}$ Tag'' events~(a, b) and ``$\Lambda_{c}\pi$ Tag'' events~(c, d) at $\sqrt{s} =$  4918.02~(a, c)  and  4950.93~MeV~(b, d). 
The blue lines are the sum of the fit functions, the red lines indicate the signal, and the green lines are the background.
The region between the red dashed lines is the $\Lambda_{c}^{+}$ signal region, and the regions between the two neighboring pink arrows are the $\Lambda_{c}^{+}$ sideband regions.
} \label{fig1}
\end{center}
\end{figure}
	%%%%%%%%%%%%%%%%%%%%%%%%%%%%%%%%%%%%%%%%
To determine the Born cross section  ($\sigma$) of the signal process, an unbinned maximum likelihood fit is performed simultaneously on the two types of tagged events at each energy point. Explicitly, the fit comprises two categories: 
\begin{itemize}
\item[(a)] For the  ``$\Lambda_{c}$ Tag'' events, the single variable  $RM(\Lambda_{c}^{+})+M(\Lambda_{c}^{+})-m(\Lambda_{c}^{+})$ is used in the fit,  where $RM(\Lambda_{c}^{+})$ is the recoiling mass  against the  reconstructed   $\Lambda_{c}^{+}$ in the c.m. system, $M(\Lambda_{c}^{+})$ is the mass of the reconstructed $\Lambda_{c}^{+}$, and m($\Lambda_{c}^{+}$) is  the known $\Lambda_{c}^{+}$ mass from  the PDG;
\item[(b)] For the  ``$\Lambda_{c}\pi$ Tag'' events,  the fit is two-dimensional, using the two variables  $RM(\Lambda_{c}^{+}\pi^{\pm})+M(\Lambda_{c}^{+})-m(\Lambda_{c}^{+})$ and  $ RM(\pi^{\pm})$,  where  $RM(\Lambda_{c}^{+}\pi^{\pm})$  and $RM(\pi^{\pm})$ are the recoiling mass against the selected $\Lambda_{c}^{+}\pi^{\pm}$ system and the $\pi^{\pm}$, respectively. 
\end{itemize}

The background components are categorized into two classes: 
\begin{itemize}
\item[(a)] Contributions from $\ee \to q\bar{q}$ processes. These are described using the sideband regions of the $M(pK^{-}\pi^{+})$ distribution  in the data, as shown in Fig.~\ref{fig1}.
The signal region is the same  as the $\Lambda_{c}^{+}$ mass window,  and the sideband regions are  [2.190, 2.250] and [2.320, 2.380] GeV/$c^{2}$. 
 To determine the number of background events, we fit this distribution.
A Gaussian function, along with a second-order Chebyshev polynomial, is used to fit the  $M(pK^{-}\pi^{+})$ distribution in data, representing the signal and background contributions, respectively.

\item[(b)] Contributions from  the processes $\ee \to \Sigma_{c}\bar{\Sigma}_{c}$, $\ee \to \Lambda_{c} \bar{\Sigma}_{c} \pi (\Sigma_{c}=\Sigma^{0}_{c}, \Sigma^{+}_{c}, \rm{and}~\Sigma^{++}_{c})$, $\ee \to \Lambda_{c}\bar{\Lambda}_{c}(2595)$, and $\ee \to \Lambda_{c}\bar{\Lambda}_{c}(2625)$.   These share the same final state ($\pi^{+}\pi^{-}\lamcplamcm$) as the signal process. 
For $\ee \to \Sigma_{c}\bar{\Sigma}_{c}$ and $\ee \to \Lambda_{c}\bar{\Sigma}_{c}\pi$, the Born cross sections  are unknown and are therefore treated as free parameters in the fit.
The Born cross sections for $\ee \to \Lambda_{c}\bar{\Lambda}_{c}(2595)$ and $\ee \to \Lambda_{c}\bar{\Lambda}_{c}(2625)$ are taken from Ref.~\cite{fjh}.
\end{itemize}
The shapes of  background (b) and the signal process are extracted from MC samples.

The Born cross section  for the signal is calculated as follows:
\begin{align}
\label{1}
\begin{split}
 \sigma =\frac{ N^{\mathrm{obs}}_{i} } {  \mathcal{L}_{\mathrm{int}} \cdot f_{\mathrm{VP}} \cdot f_{\mathrm{ISR}} \cdot  \mathcal{B} \cdot \epsilon_{i}},
\end{split}
\end{align}
where $i=1,2$ label the ``$\Lambda_{c}$ Tag'' and  ``$\Lambda_{c}\pi$ Tag'' modes; $N_{i}^{\rm{obs}}$ represents the fit yield in each mode; and $\epsilon_{i}$ denotes the event selection efficiencies.
For all signal MC samples: at $\sqrt{s} =$ 4918.02~MeV,  $\epsilon_{1}$ ranges from 17.8\% to 18.4\% and  $\epsilon_{2}$ ranges  from 32.4\% to 34.3\%; at $\sqrt{s} =$ 4950.93~MeV, $\epsilon_{1}$ ranges from 13.3\% to 14.7\% and $\epsilon_{2}$ ranges from 38.0\% to 42.1\%.
$\mathcal{B}$ is the BF of $\lambdacp \rightarrow pK^{-}\pi^{+}$, with a value of (6.28$\pm$0.32)\% quoted from the PDG. 
$\mathcal{L}_{\mathrm{int}}$ represents the integrated luminosity.
The vacuum polarization factor, $f_{\mathrm{VP}}$, is calculated to be 1.056  at the two c.m. energies~\cite{fvp-1,fvp-2}.
The  ISR correction factor $f_{\mathrm{ISR}}$~\cite{fisr} is obtained using the generator {\sc kkmc}, yielding values of  1.42 at $\sqrt{s} =$ 4918.02~MeV for all $H_c$ MC samples  and 1.55 at $\sqrt{s} =$ 4950.93~MeV. The input Born cross sections are assumed to be the same as $\ee \to \Lambda_{c} \bar{\Lambda}_{c}$~\cite{shapeLambadc}.
For the $\ee \to \Lambda_{c} \bar{\Lambda}^{\ast}_{c} $ process, the $f_{\mathrm{ISR}}$ value is taken from Ref.~\cite{fjh}.
Additionally, for the processes $\ee \to \Sigma_{c}\bar{\Sigma}_{c}$ and $\ee \to \Lambda_{c}\bar{\Sigma}_{c} \pi$ at the two c.m. energies, the $f_{\mathrm{ISR}}$ values are 0.90, assuming a flat Born cross section line shape.

The statistical significance of the $H_{c}$ is estimated by comparing the likelihoods of the fits with and without the inclusion of the signal component.
All $H_{c}$ hypotheses except one have a significance below 3.0$\sigma$. For the $H_{c}$ at a mass of 4720~MeV/$c^{2}$ and width of 5~MeV at $\sqrt{s}=4918.02$~MeV, the significance is 3.0$\sigma$. 
After considering the look elsewhere effect~\cite{else} and systematic uncertainty, the significance is less than 2.1$\sigma$.
The fit results for this $H_{c}$ are presented in Fig.~\ref{fig2} for both c.m. energies. 
 Additionally,  the $H_{c}$ with mass 4720~MeV/$c^{2}$ serves as a representative case in the subsequent discussion on systematic uncertainties.
In  Section~\ref{sec6}, we determine the upper limits of  $\sigma(\ee \to \pi^{+}  H_c^{-} + c.c.) \times \mathcal{B}(H_c^{-} \rightarrow  \pi^{-}\lamcplamcm )$ for all $H_{c}$ hypotheses, taking systematic uncertainties into account.

\begin{figure*}[htbp]
\begin{center}
\includegraphics[width=1\textwidth]{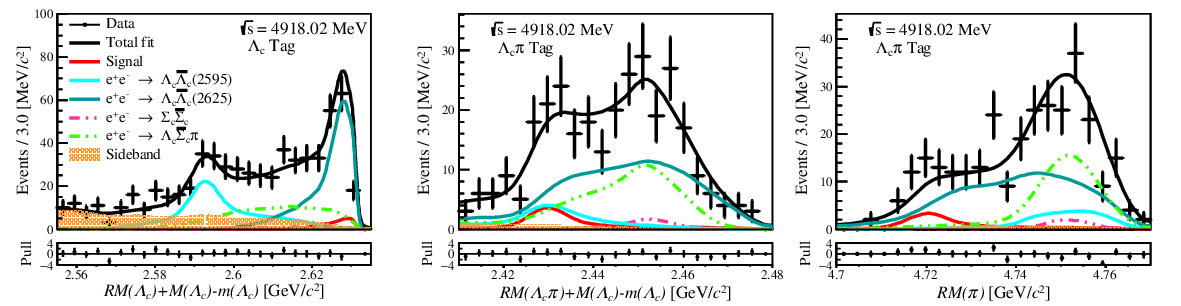}
\includegraphics[width=1\textwidth]{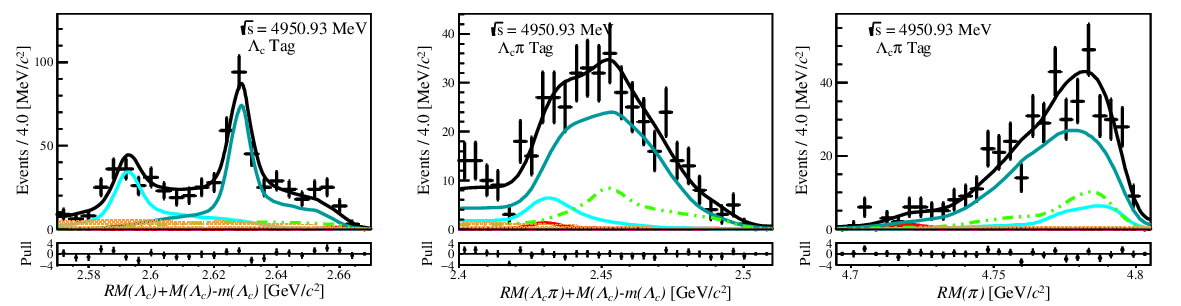}
\caption{Simultaneous fit results at $\sqrt{s} =$ 4918.02~MeV~(top) and  4950.93~MeV~(bottom). The first column is for the distribution of $RM(\Lambda_{c}^{+})+M(\Lambda_{c}^{+})-m(\Lambda_{c}^{+})$. The second and third columns display  the one-dimensional projections of  $RM(\Lambda_{c}^{+}\pi^{\pm})+M(\Lambda_{c}^{+})-m(\Lambda_{c}^{+})$ versus  $ RM(\pi^{\pm})$, respectively. 
The dots with error bars are data, the black solid curves correspond to the best fit results, the red solid lines show the signal shapes, the cyan solid lines represent the contribution from $\ee \to \Lambda_{c}\bar{\Lambda}_{c}(2595)$, the blue solid lines represent the contribution from $\ee \to \Lambda_{c}\bar{\Lambda}_{c}(2625)$, the pink dashed lines are the contribution from $\ee \to \Sigma_{c}\bar{\Sigma}_{c}$, the green dashed lines are the contribution from $\ee \to \Lambda_{c}\bar{\Sigma}_{c} \pi$, 
the shaded histograms show the $\Lambda_{c}^{+}$ sideband contributions from each tag's own sideband region.} \label{fig2}
\end{center}
\end{figure*}

	%%%%%%%%%%%%%%%%%%%%%%%%%%%%%%%%%%%%%%%%
	%      5. Systematic Uncertainty
	%%%%%%%%%%%%%%%%%%%%%%%%%%%%%%%%%%%%%%%
\section{Systematic Uncertainties}\label{sec5}
In the cross section measurements, the  systematic uncertainties are categorized into multiplicative (i-vi) and additive terms (vii-ix), and are assumed to be independent. 
In Section~\ref{sec6}, we present the upper limit on the signal Born cross section using the Bayesian approach~\cite{baye}. This method accounts for systematic uncertainties in two stages. Multiplicative systematic uncertainties are incorporated by convolving the likelihood distribution with a Gaussian function. For the additive terms, the largest upper limit of the signal Born cross section from different fit conditions is selected. The summation of the multiplicative terms is shown in Table~\ref{table1}. For the additive terms, we describe the methods used to estimate them below.

\begin{itemize}

\item[(i)]{\bf Luminosity:}
The integrated luminosity for each data sample is measured using  Bhabha events,  with an uncertainty of 0.7\%~\cite{cms}.

\item[(ii)]{\bf MC statistics:}
The systematic uncertainties arising from the finite MC sample sizes are calculated by $\sqrt{(1-\epsilon)/\epsilon}/\sqrt{N}$, where $\epsilon$ is the reconstruction efficiency after all the event selection, and $N$ is the number of generated events. 
At $\sqrt{s} =$ 4918.02~MeV, the uncertainties  are  0.9\% for the ``$\Lambda_{c}$ Tag''  and  0.6\% for the ``$\Lambda_{c} \pi$ Tag''. At $\sqrt{s} =$  4950.93~MeV,  the uncertainties are  1.0\%  for the  ``$\Lambda_{c}$ Tag''  and  0.6\% for the ``$\Lambda_{c} \pi$ Tag''. These values remain consistent across the $H_{c}$ widths of 5, 10, and 20~MeV.

\item[(iii)]{\bf Input BF:}
The systematic uncertainty from the BF of $\lambdacp \rightarrow pK^{-}\pi^{+}$  is  5.0\%.

\item[(iv)]{\bf Tracking and PID:}
 The systematic uncertainty due to tracking or PID is assigned as 1.0\%  for each charged track~\cite{pid_sys}.
They are applied for three tracks in the ``$\Lambda_{c}$ Tag'' ($pK^{-}\pi^{+}$) and four tracks in the  ``$\Lambda_{c} \pi$ Tag'' ($pK^{-}\pi^{+} \pi^{+}$).
Therefore, the combined tracking and PID   systematic uncertainties for the ``$\Lambda_{c}$ Tag'' and ``$\Lambda_{c} \pi$ Tag'' are 4.3\% and 5.7\%, respectively.

\item[(v)]{\bf $\Lambda_{c}^{+}$  vertex fit:}
This systematic uncertainty arises from  the difference in the $pK^{-}\pi^{+}$ vertex fit efficiencies between the MC simulation and data.
Using a  control sample of  $e^{+}e^{-} \to \Lambda_{c} \bar{\Lambda}_{c}$, we find a difference of 2.0\%, which is taken as the systematic uncertainty.

\item[(vi)]{\bf Mass window of $\Lambda_{c}^{+}$:} 
We use either the direct signal shape or a signal shape convolved with a Gaussian function to fit the data. The proportions of signal events within the  $\Lambda_{c}^{+}$ mass window serve as the resolution for the MC or data, respectively. The difference between these two ratios is regarded as the uncertainty.  For $\sqrt{s} =$ 4918.02~MeV, the differences for $H_{c}$ widths of  5, 10, and 20~MeV  are  2.2\%, 2.1\%, and  2.2\%, respectively.  For $\sqrt{s} =$  4950.93~MeV,  the differences are all 1.5\%.

\item[(vii)]{\bf Fit range:}
The systematic uncertainty due to the fit range is tested using a Barlow test method~\cite{Barlow}.
 By varying the fit ranges, we find that the change in the signal Born cross section is much less than the statistical uncertainty, so  the systematic uncertainties are negligible.

\item[(viii)]{\bf Signal shape:}
The signal MC samples are generated with the Born cross section line shape of $\ee \to \Lambda_{c} \bar{\Lambda}_{c} $, as no measurement has been done before. To assess this uncertainty, we vary the line shape to a flat line shape and  generate signal MC samples.

\item[(ix)]{\bf Background:}
The two types of  background (a) and (b)  introduce four sources of systematic uncertainties. 

(1) To estimate the uncertainties in the $M(pK^{-}\pi^{+})$ sideband regions from data, we modify  these regions to use those derived from the inclusive $\ee \to q\bar{q}$  MC sample.

(2) For the processes $\ee \to \Lambda_{c} \bar{\Sigma}_{c} \pi$ and $\ee \to \Sigma_{c} \bar{\Sigma}_{c}$, where the cross section line shape is unknown, we assume a flat line shape. To estimate the systematic uncertainty, we alter the line shape from flat to that of $\ee \to \Lambda_{c} \bar{\Lambda}_{c}$.

(3) We replace the two original background shapes of $\ee \to \Lambda_{c} \bar{\Sigma}_{c} \pi$ and $\ee \to \Sigma_{c} \bar{\Sigma}_{c}$ with  a single background shape of $\ee \to \Lambda_{c} \bar{\Sigma}_{c} \pi$. 

(4) For the process $\ee \to \Lambda_{c} \bar{\Lambda}^{\ast}_{c}$, we use the central Born cross section values from Ref.~\cite{fjh}, which has considered the uncertainty in BFs of $\bar{\Lambda}^{\ast}_{c}$. 
The Born cross sections of this process vary within their error ranges during fitting.

\end{itemize}

\begin{table}[htbp]
\caption{The multiplicative systematic uncertainties (in \%) for the $H_{c}$ with a mass of 4720~MeV/$c^{2}$ and the widths 5, 10, and 20~MeV at two c.m. energies. The number ``$\ast$/$\ast$'' refers to the ``$\Lambda_{c}$ Tag'' and the ``$\Lambda_{c}\pi$ Tag'', respectively.}
\centering
\begin{tabular}{l| ccc |ccc}
\hline  \hline 

c.m. energy (MeV) & \multicolumn{3}{c|}{4918.02} &\multicolumn{3}{c}{4950.93}    \\ \hline

$H_{c}$ width  (MeV) &5  &10  & 20 &5  &10 & 20    \\ \hline \hline
Luminosity                     & \multicolumn{6}{c}{0.7}    \\ 
MC statistics                  & \multicolumn{3}{c|}{0.9/0.6}   & \multicolumn{3}{c}{1.0/0.6} \\ 
Input BF                       & \multicolumn{6}{c}{5.0}  \\ 
Tracking and PID              & \multicolumn{6}{c}{4.3/5.7}    \\ 
$\Lambda_{c}^{+}$  vertex fit      & \multicolumn{6}{c}{2.0}   \\ 
Mass window of $\Lambda_{c}^{+}$   & 2.2  & 2.1  & 2.2  & 1.5   & 1.5  & 1.5   \\ \hline
\textbf{Sum} for ``$\Lambda_{c}$ Tag''      & 7.3  & 7.3  & 7.3  & 7.1  & 7.1   & 7.1   \\
\textbf{Sum} for ``$\Lambda_{c}\pi$ Tag''  & 8.2  &  8.2  &  8.2  & 8.1   & 8.1  & 8.1    \\ \hline \hline 

\end{tabular}\label{table1}
\end{table}

\begin{figure}[]
\begin{center}
\begin{minipage}[t]{1.0\linewidth}
\includegraphics[width=0.8\textwidth]{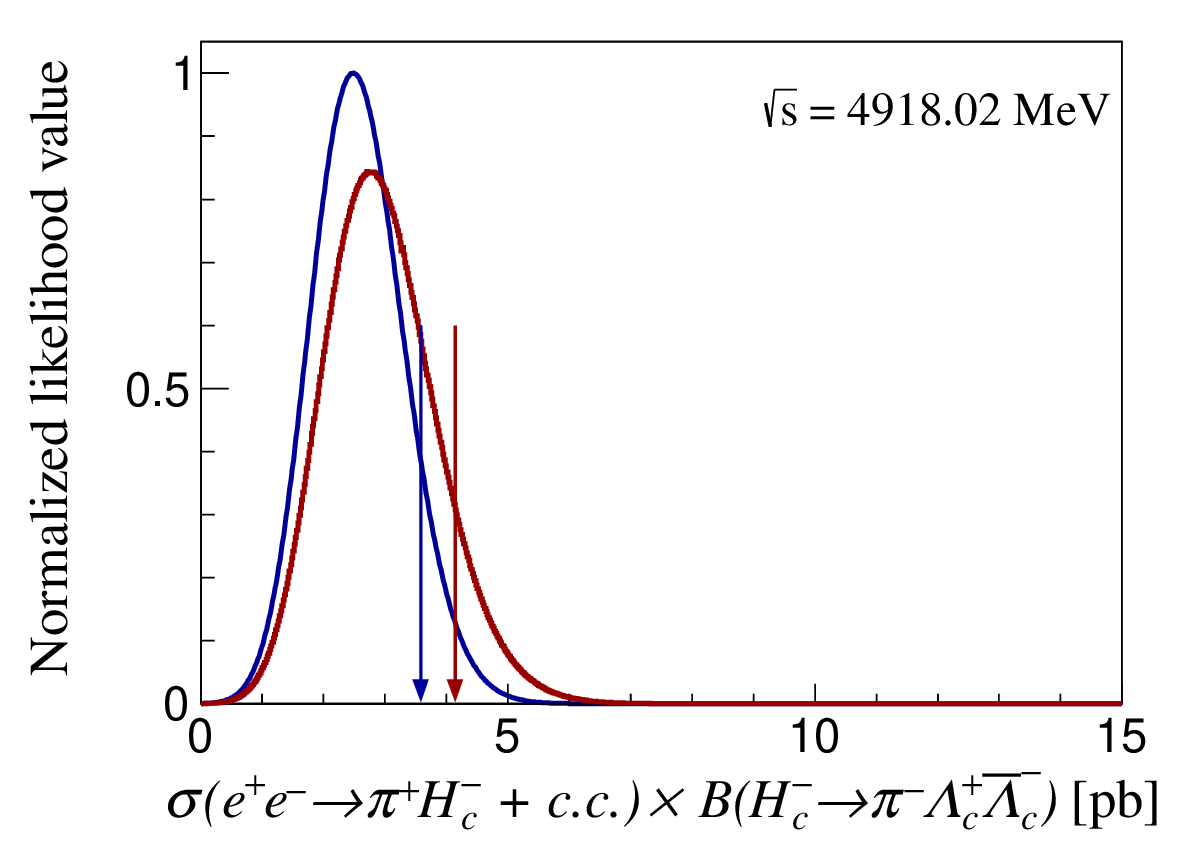}
\includegraphics[width=0.8\textwidth]{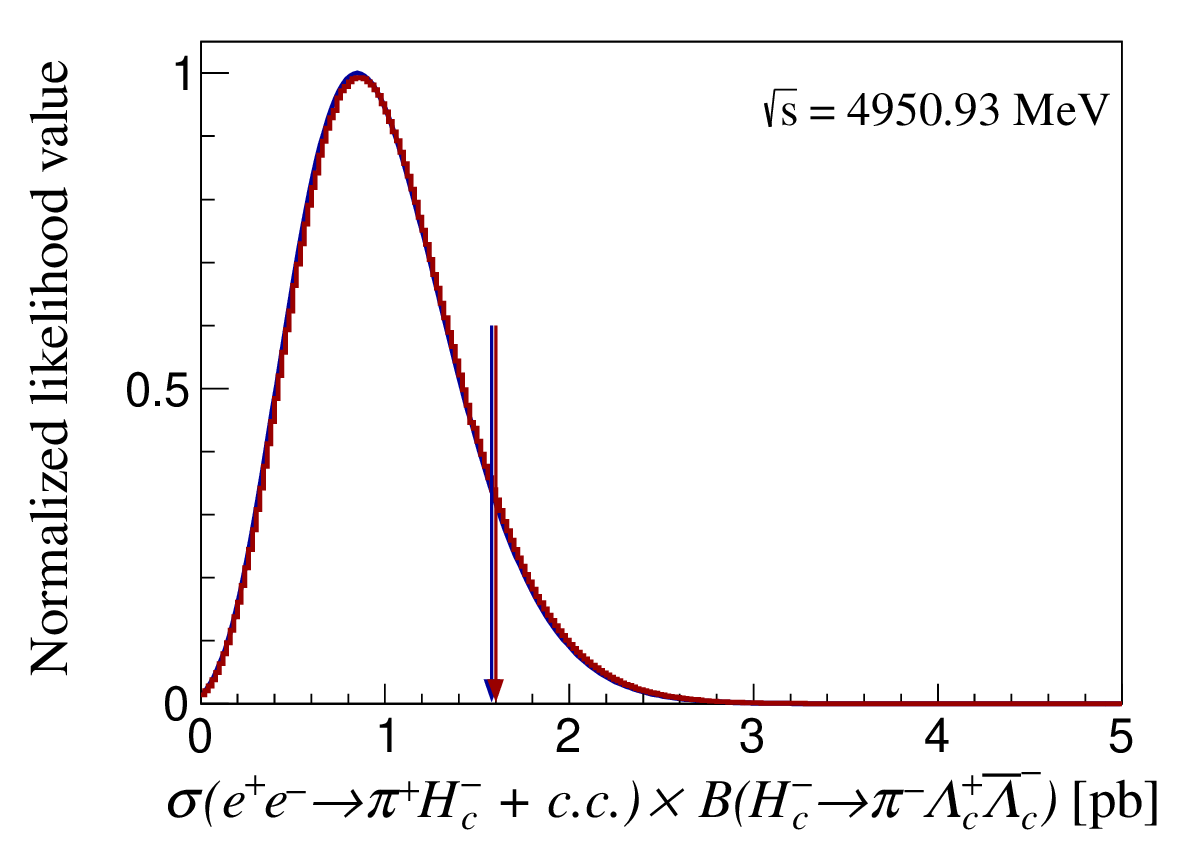}
\end{minipage}
\caption{Distributions of  the likelihood versus $\sigma(\ee \to \pi^{+}  H_c^{-} + c.c.) \times \mathcal{B}(H_c^{-} \rightarrow  \pi^{-}\lamcplamcm  )$ at $\sqrt{s} =$ 4918.02~MeV~(top) and  4950.93~MeV~(bottom).  The blue and red histograms denote those without and with the incorporation of systematic uncertainties.}\label{fig4}
\end{center}
\end{figure}

\begin{figure}[]
\begin{center}
\begin{minipage}[t]{1.0\linewidth}
\includegraphics[width=0.98\textwidth]{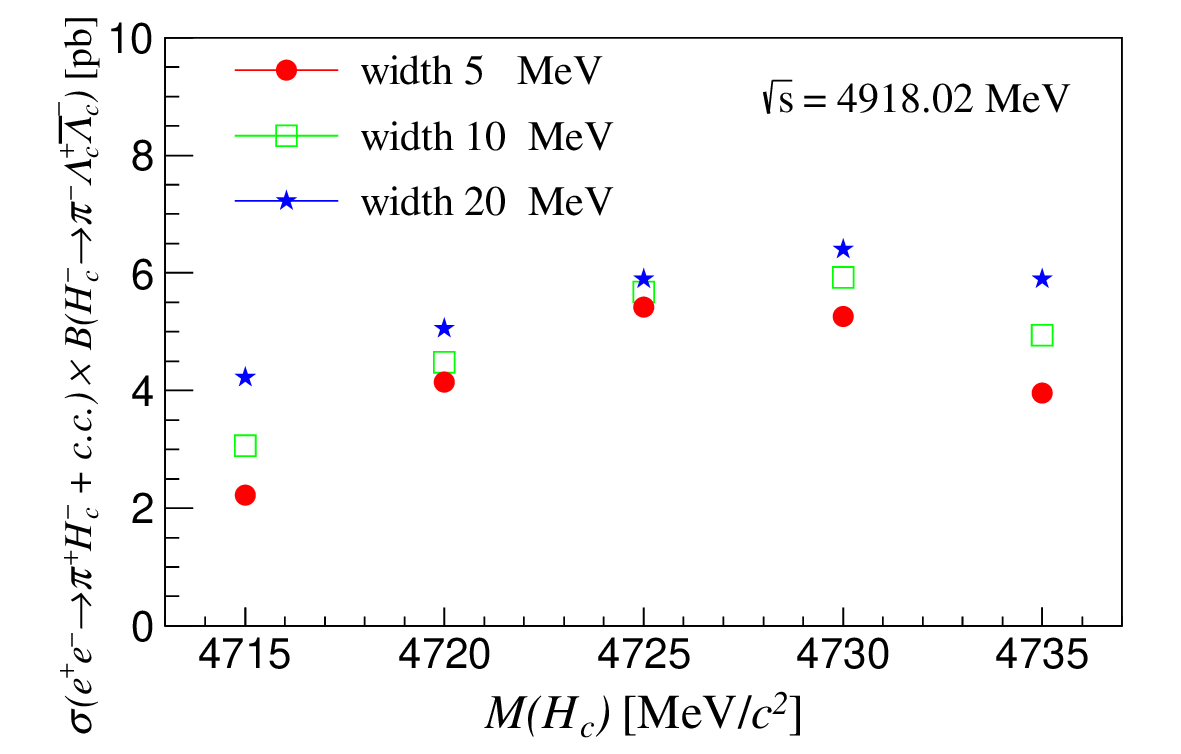}
\includegraphics[width=0.98\textwidth]{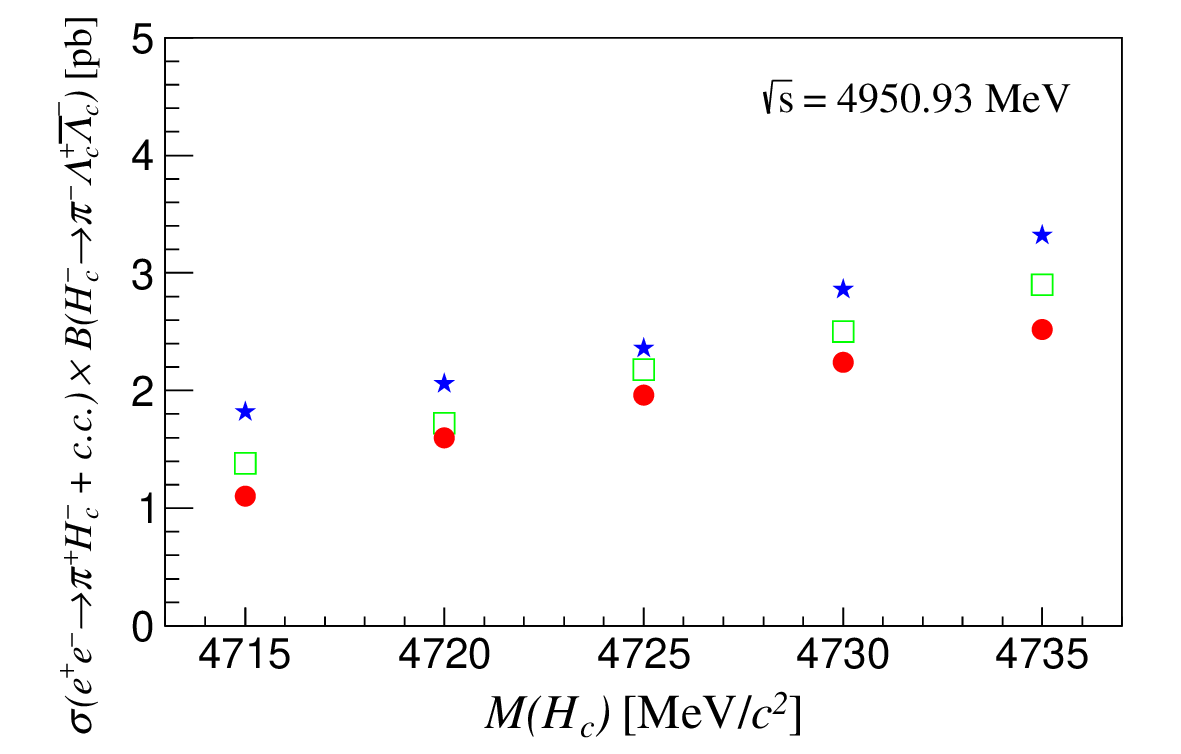}
\end{minipage}
\caption{Upper limits on  $\sigma(\ee \to \pi^{+}  H_c^{-} + c.c.) \times \mathcal{B}(H_c^{-} \rightarrow   \pi^{-}\lamcplamcm)$ at $\sqrt{s} =$ 4918.02~MeV~(top) and 4950.93~MeV~(bottom) for all signal $H_{c}$.  }\label{figb}
\end{center}
\end{figure}

        %%%%%%%%%%%%%%%%%%%%%%%%%%%%%%%%%%%%%%%%
	%         6. Result
	%%%%%%%%%%%%%%%%%%%%%%%%%%%%%%%%%%%%%%%%
\section{Result}\label{sec6}
Due to its low statistical significance, the upper limit on  $\sigma(\ee \to \pi^{+}  H_c^{-} + c.c.) \times \mathcal{B}(H_c^{-} \rightarrow  \pi^{-}\lamcplamcm )$ is set at a 90\% confidence level (C.L.).
To account for systematic uncertainties when determining the upper limit, we define a new likelihood distribution~\cite{likehood} based on the following equation:
\begin{align}
\begin{split}\label{eqUper}
    \tilde{L}(\sigma)=\int L( \frac{\epsilon  }{\epsilon_{0} }\sigma  )  \frac{1}{\sqrt{2\pi  \sigma_{\epsilon}^2}}\exp{[-\frac{(\epsilon-\epsilon_{0})^2}{2 \sigma_{\epsilon}^2}]} \mathrm{d}\epsilon,
\end{split}
\end{align}
where $L$ is the  original likelihood distribution,   $\epsilon_{0}$ represents the nominal selection efficiency, and $\sigma_{\epsilon}$ denotes the multiplicative systematic uncertainties.
Subsequently, the maximum upper limit of the signal Born cross section is determined by taking into account the conditions of various additive systematic terms from different fits.
In the simultaneous fit, two likelihoods from the ``$\Lambda_{c}$ Tag'' and ``$\Lambda_{c}\pi$ Tag''  are combined multiplicatively to construct the total likelihood. 
Therefore, the two likelihoods are adjusted using Eq.~(\ref{eqUper}) and then multiplied to derive the total $\tilde{L}$,  which is then used to determine the upper limit.
As an example, Fig.~\ref{fig4} displays the distributions of  the likelihood versus $\sigma(\ee \to \pi^{+}  H_c^{-} + c.c.) \times \mathcal{B}(H_c^{-} \rightarrow  \pi^{-}\lamcplamcm)$  for the $H_{c}$ with a mass of 4720~MeV/$c^{2}$ and a width of 5~MeV at two c.m. energies. 
The upper limits on $\sigma(\ee \to \pi^{+} H_c^{-} + c.c.) \times \mathcal{B}(H_c^{-} \rightarrow \pi^{-}\lamcplamcm )$ for all $H_{c}$ are depicted in Fig.~\ref{figb}, with systematic uncertainties considered.

	%%%%%%%%%%%%%%%%%%%%%%%%%%%%%%%%%%%%%%%%
	%         7. Summary                   
	%%%%%%%%%%%%%%%%%%%%%%%%%%%%%%%%%%%%%%%%
\section{Summary}
In this work, we search for a possible $\Lambda_{c} \bar{{\Sigma}}_{c}$ bound state, denoted $H_c$, using 208.11 and 160.37 $\ipb$ of $\ee$ annihilation data at $\sqrt{s} =$ 4918.02 and 4950.93~MeV, respectively, collected with the BESIII detector.
No evidence for this state is observed.
We set upper limits on  $\sigma(\ee \to \pi^{+}  H_c^{-} + c.c ) \times \mathcal{B}(H_c^{-} \rightarrow  \pi^{-}\lamcplamcm  )$ at a 90\% C.L., for the $H_{c}$ assuming masses of 4715, 4720, 4725, 4730, and 4735~MeV/$c^{2}$, and widths of 5, 10, and 20~MeV.
The final upper limits, as shown in Fig.~\ref{figb},  range from 1.1~pb to 6.4~pb.
This result represents the first search for a $\Lambda_{c}\bar{\Sigma}_{c}$ bound state. We note that the background level is high and the data statistics are limited. In the future, the upgraded BEPCII~\cite{BESIII:2020nme} and a next-generation super $\tau$-charm factory~\cite{stcf} could provide more data to  identify  this signal process.

	%%%%%%%%%%%%%%%%%%%%%%%%%%%%%%%%%%%%%%%%
	%         8. ACKNOWLEDGMENTS
	%%%%%%%%%%%%%%%%%%%%%%%%%%%%%%%%%%%%%%%%
\section*{ACKNOWLEDGMENTS}
The BESIII Collaboration thanks the staff of BEPCII (https://cstr.cn/31109.02.BEPC) and the IHEP computing center for their strong support. This work is supported in part by National Key R\&D Program of China under Contracts Nos. 2023YFA1606000, 2023YFA1606704; National Natural Science Foundation of China (NSFC) under Contracts Nos. 11635010, 11935015, 11935016, 11935018, 12025502, 12035009, 12035013, 12061131003, 12192260, 12192261, 12192262, 12192263, 12192264, 12192265, 12221005, 12225509, 12235017, 12361141819; the Chinese Academy of Sciences (CAS) Large-Scale Scientific Facility Program; CAS under Contract No. YSBR-101; 100 Talents Program of CAS; The Institute of Nuclear and Particle Physics (INPAC) and Shanghai Key Laboratory for Particle Physics and Cosmology; ERC under Contract No. 758462; German Research Foundation DFG under Contract No. FOR5327; Istituto Nazionale di Fisica Nucleare, Italy; Knut and Alice Wallenberg Foundation under Contracts Nos. 2021.0174, 2021.0299; Ministry of Development of Turkey under Contract No. DPT2006K-120470; National Research Foundation of Korea under Contract No. NRF-2022R1A2C1092335; National Science and Technology fund of Mongolia; Polish National Science Centre under Contract No. 2024/53/B/ST2/00975; STFC (United Kingdom); Swedish Research Council under Contract No. 2019.04595; U. S. Department of Energy under Contract No. DE-FG02-05ER41374


\begin{thebibliography}{99}


\bibitem{xyz-ycz}
 N.~Brambilla, S.~Eidelman, C.~Hanhart, A.~Nefediev, C.~P.~Shen, C.~E.~Thomas, A.~Vairo and C.~Z.~Yuan,
\href{https://www.sciencedirect.com/science/article/pii/S0370157320301915?via%3Dihub}{Phys. Rept. \textbf{873}, 1 (2020)}.

% %\cite{Brambilla:2019esw}
% \bibitem{Brambilla:2019esw}
% N.~Brambilla, S.~Eidelman, C.~Hanhart, A.~Nefediev, C.~P.~Shen, C.~E.~Thomas, A.~Vairo and C.~Z.~Yuan,
% %``The $XYZ$ states: experimental and theoretical status and perspectives,''
% Phys. Rept. \textbf{873}, 1-154 (2020)
% doi:10.1016/j.physrep.2020.05.001
% [arXiv:1907.07583 [hep-ex]].
% %915 citations counted in INSPIRE as of 16 Jun 2025


\bibitem{xyz1}
S.~K.~Choi {\it et al.} (BELLE Collaboration),
\href{https://journals.aps.org/prl/abstract/10.1103/PhysRevLett.91.262001}{Phys. Rev. Lett, \textbf{91}, 262001 (2003)}.

% %\cite{Belle:2003nnu}
% \bibitem{Belle:2003nnu}
% S.~K.~Choi \textit{et al.} [Belle],
% %``Observation of a narrow charmonium-like state in exclusive $B^\pm \to K^\pm \pi^+ \pi^- J/\psi$ decays,''
% Phys. Rev. Lett. \textbf{91}, 262001 (2003)
% doi:10.1103/PhysRevLett.91.262001
% [arXiv:hep-ex/0309032 [hep-ex]].
% %2752 citations counted in INSPIRE as of 29 May 2025

\bibitem{xyz2}
M.~Ablikim {\it et al.} (BESIII Collaboration),
\href{https://journals.aps.org/prl/abstract/10.1103/PhysRevLett.110.252001}{Phys. Rev. Lett, \textbf{110}, 252001 (2013)}.

% %\cite{BESIII:2013ris}
% \bibitem{BESIII:2013ris}
% M.~Ablikim \textit{et al.} [BESIII],
% %``Observation of a Charged Charmoniumlike Structure in $e^+e^- \to \pi^+\pi^- J/\psi$ at $\sqrt{s}$ =4.26 GeV,''
% Phys. Rev. Lett. \textbf{110}, 252001 (2013)
% doi:10.1103/PhysRevLett.110.252001
% [arXiv:1303.5949 [hep-ex]].
% %1192 citations counted in INSPIRE as of 29 May 2025

\bibitem{molecular}
F.~K.~Guo, C.~Hanhart, U.~G.~Mei\ss{}ner, Q.~Wang, Q.~Zhao and B.~S.~Zou,
\href{https://journals.aps.org/rmp/abstract/10.1103/RevModPhys.90.015004}{Rev. Mod. Phys, \textbf{90}, 015004 (2018)}.

% %\cite{Guo:2017jvc}
% \bibitem{Guo:2017jvc}
% F.~K.~Guo, C.~Hanhart, U.~G.~Mei\ss{}ner, Q.~Wang, Q.~Zhao and B.~S.~Zou,
% %``Hadronic molecules,''
% Rev. Mod. Phys. \textbf{90}, no.1, 015004 (2018)
% [erratum: Rev. Mod. Phys. \textbf{94}, no.2, 029901 (2022)]
% doi:10.1103/RevModPhys.90.015004
% [arXiv:1705.00141 [hep-ph]].
% %1356 citations counted in INSPIRE as of 29 May 2025


\bibitem{dxk}
X.~K.~Dong, F.~K.~Guo and B.~S.~Zou,
\href{https://pip.nju.edu.cn/CN/10.13725/j.cnki.pip.2021.02.001}{Prog. Phys.  \textbf{41}, 65-93 (2021)}.

% %\cite{Dong:2021juy}
% \bibitem{Dong:2021juy}
% X.~K.~Dong, F.~K.~Guo and B.~S.~Zou,
% %``A survey of heavy-antiheavy hadronic molecules,''
% Progr. Phys. \textbf{41}, 65-93 (2021)
% doi:10.13725/j.cnki.pip.2021.02.001
% [arXiv:2101.01021 [hep-ph]].
% %176 citations counted in INSPIRE as of 29 May 2025

\bibitem{cms}
M.~Ablikim {\it et al.} (BESIII Collaboration),
\href{https://iopscience.iop.org/article/10.1088/1674-1137/ac84cc}{Chin. Phys. C \textbf{46}, 113003 (2022)}.

% %\cite{BESIII:2022ulv}
% \bibitem{BESIII:2022ulv}
% M.~Ablikim \textit{et al.} [BESIII],
% %``Luminosities and energies of e $^{+}$ e $^{−}$ collision data taken between =4.61 GeV and 4.95 GeV at BESIII*,''
% Chin. Phys. C \textbf{46}, no.11, 113003 (2022)
% doi:10.1088/1674-1137/ac84cc
% [arXiv:2205.04809 [hep-ex]].
% %83 citations counted in INSPIRE as of 29 May 2025


\bibitem{Ablikim:2009aa} M. Ablikim {\it et al.} (BESIII Collaboration), \href{https://doi.org/10.1016/j.nima.2009.12.050}{Nucl. Instrum. Meth. A {\bf 614}, 345 (2010)}.

% %\cite{BESIII:2009fln}
% \bibitem{BESIII:2009fln}
% M.~Ablikim \textit{et al.} [BESIII],
% %``Design and Construction of the BESIII Detector,''
% Nucl. Instrum. Meth. A \textbf{614}, 345-399 (2010)
% doi:10.1016/j.nima.2009.12.050
% [arXiv:0911.4960 [physics.ins-det]].
% %1365 citations counted in INSPIRE as of 29 May 2025

\bibitem{CXYu_bes3} C.~H.~Yu {\it et al.}, in \href{https://accelconf.web.cern.ch/ipac2016/doi/JACoW-IPAC2016-TUYA01.html} {7th International Particle Accelerator Conference} (2016) p. TUYA01.


\bibitem{tof_a} X.~Li \textit{et al.}, \href{https://link.springer.com/article/10.1007\%2Fs41605-017-0014-2}{Radiat. Detect. Technol. Meth. {\bf 1}, 13 (2017)}.

% %\cite{Li:2017jpg}
% \bibitem{Li:2017jpg}
% X.~Li, Y.~Sun, C.~Li, Z.~Liu, Y.~Heng, M.~Shao, X.~Wang, Z.~Wu, P.~Cao and M.~Chen, \textit{et al.}
% %``Study of MRPC technology for BESIII endcap-TOF upgrade,''
% Radiat. Detect. Technol. Methods \textbf{1}, 13 (2017)
% doi:10.1007/s41605-017-0014-2
% %255 citations counted in INSPIRE as of 29 May 2025

\bibitem{tof_b}  Y.~X.~Guo \textit{et al.}, \href{https://link.springer.com/article/10.1007\%2Fs41605-017-0012-4}{Radiat. Detect. Technol. Meth. {\bf 1}, 15 (2017)}.


% %\cite{Guo:2017sjt}
% \bibitem{Guo:2017sjt}
% Y.~X.~Guo, S.~S.~Sun, F.~F.~An, R.~X.~Yang, M.~Zhou, Z.~Wu, H.~L.~Dai, Y.~K.~Heng, C.~Li and Z.~Y.~Deng, \textit{et al.}
% %``The study of time calibration for upgraded end cap TOF of BESIII,''
% Radiat. Detect. Technol. Methods \textbf{1}, 15 (2017)
% doi:10.1007/s41605-017-0012-4
% %249 citations counted in INSPIRE as of 29 May 2025

\bibitem{tof_c}P.~Cao \textit{et al.}, \href{https://www.sciencedirect.com/science/article/pii/S0168900219314068?via\%3Dihub}{Nucl. Instrum. Meth. A {\bf 953}, 163053 (2020)}.


% %\cite{Cao:2020ibk}
% \bibitem{Cao:2020ibk}
% P.~Cao, H.~F.~Chen, M.~M.~Chen, H.~L.~Dai, Y.~K.~Heng, X.~L.~Ji, X.~S.~Jiang, C.~Li, X.~Li and S.~B.~Liu, \textit{et al.}
% %``Design and construction of the new BESIII endcap Time-of-Flight system with MRPC Technology,''
% Nucl. Instrum. Meth. A \textbf{953}, 163053 (2020)
% doi:10.1016/j.nima.2019.163053
% %207 citations counted in INSPIRE as of 29 May 2025


\bibitem{geant4} S.~Agostinelli \textit{et al.} (GEANT4 Collaboration), \href{https://doi.org/10.1016/S0168-9002(03)01368-8}{Nucl. Instrum. Meth. A {\bf 506}, 250 (2003)}.

% %\cite{GEANT4:2002zbu}
% \bibitem{GEANT4:2002zbu}
% S.~Agostinelli \textit{et al.} [GEANT4],
% %``GEANT4 - A Simulation Toolkit,''
% Nucl. Instrum. Meth. A \textbf{506}, 250-303 (2003)
% doi:10.1016/S0168-9002(03)01368-8
% %20343 citations counted in INSPIRE as of 29 May 2025

\bibitem{detvis} K.~X.~Huang {\it et al.}, \href{https://doi.org/10.1007/s41365-022-01133-8}{Nucl.\ Sci.\ Tech. {\bf 33}, 142 (2022)}.

% %\cite{Huang:2022wuo}
% \bibitem{Huang:2022wuo}
% K.~X.~Huang, Z.~J.~Li, Z.~Qian, J.~Zhu, H.~Y.~Li, Y.~M.~Zhang, S.~S.~Sun and Z.~Y.~You,
% %``Method for detector description transformation to Unity and application in BESIII,''
% Nucl. Sci. Tech. \textbf{33}, no.11, 142 (2022)
% doi:10.1007/s41365-022-01133-8
% [arXiv:2206.10117 [physics.ins-det]].
% %164 citations counted in INSPIRE as of 29 May 2025


\bibitem{kkmc_a} S.~Jadach, B.~F.~L.~Ward, and Z.~Was, \href{https://www.sciencedirect.com/science/article/pii/S0010465500000485?via\%3Dihub}{Comput. Phys. Commun. {\bf 130}, 260 (2000)}.

% %\cite{Jadach:1999vf}
% \bibitem{Jadach:1999vf}
% S.~Jadach, B.~F.~L.~Ward and Z.~Was,
% %``The Precision Monte Carlo event generator K K for two fermion final states in e+ e- collisions,''
% Comput. Phys. Commun. \textbf{130}, 260-325 (2000)
% doi:10.1016/S0010-4655(00)00048-5
% [arXiv:hep-ph/9912214 [hep-ph]].
% %1323 citations counted in INSPIRE as of 29 May 2025


\bibitem{evtg1} D.~J.~Lange, \href{https://www.sciencedirect.com/science/article/pii/S0168900201000894?via\%3Dihub} {Nucl. Instrum. Meth. A {\bf 462}, 152 (2001)}.

% %\cite{Lange:2001uf}
% \bibitem{Lange:2001uf}
% D.~J.~Lange,
% %``The EvtGen particle decay simulation package,''
% Nucl. Instrum. Meth. A \textbf{462}, 152-155 (2001)
% doi:10.1016/S0168-9002(01)00089-4
% %4710 citations counted in INSPIRE as of 29 May 2025

\bibitem{evtg2} R.~G.~Ping, \href{https://iopscience.iop.org/article/10.1088/1674-1137/32/8/001}{Chin. Phys. C {\bf 32}, 599 (2008)}.

% %\cite{Ping:2008zz}
% \bibitem{Ping:2008zz}
% R.~G.~Ping,
% %``Event generators at BESIII,''
% Chin. Phys. C \textbf{32}, 599 (2008)
% doi:10.1088/1674-1137/32/8/001
% %591 citations counted in INSPIRE as of 29 May 2025



\bibitem{pdg2024}
S.~Navas  {\it et al.} (Particle Data Group),  \href{https://journals.aps.org/prd/abstract/10.1103/PhysRevD.110.030001}{Phys. Rev. D {\bf 110}, 030001 (2024)}.

% %\cite{ParticleDataGroup:2024cfk}
% \bibitem{ParticleDataGroup:2024cfk}
% S.~Navas \textit{et al.} [Particle Data Group],
% %``Review of particle physics,''
% Phys. Rev. D \textbf{110}, no.3, 030001 (2024)
% doi:10.1103/PhysRevD.110.030001
% %1723 citations counted in INSPIRE as of 29 May 2025

\bibitem{lundcharm_a} J.~C.~Chen, G.~S.~Huang, X.~R.~Qi, D.~H.~Zhang, and Y.~S.~Zhu, \href{https://journals.aps.org/prd/abstract/10.1103/PhysRevD.62.034003}{Phys. Rev. D {\bf 62}, 034003 (2000)}.

% %\cite{Chen:2000tv}
% \bibitem{Chen:2000tv}
% J.~C.~Chen, G.~S.~Huang, X.~R.~Qi, D.~H.~Zhang and Y.~S.~Zhu,
% %``Event generator for J / psi and psi (2S) decay,''
% Phys. Rev. D \textbf{62}, 034003 (2000)
% doi:10.1103/PhysRevD.62.034003
% %621 citations counted in INSPIRE as of 29 May 2025

\bibitem{lundcharm_b}  R.~L.~Yang, R.~G.~Ping, and H.~Chen, \href{https://iopscience.iop.org/article/10.1088/0256-307X/31/6/061301}{Chin. Phys. Lett. {\bf31}, 061301 (2014)}.

% %\cite{Yang:2014vra}
% \bibitem{Yang:2014vra}
% R.~L.~Yang, R.~G.~Ping and H.~Chen,
% %``Tuning and Validation of the Lundcharm Model with $J/\psi$ Decays,''
% Chin. Phys. Lett. \textbf{31}, 061301 (2014)
% doi:10.1088/0256-307X/31/6/061301
% %309 citations counted in INSPIRE as of 29 May 2025


\bibitem{photos} E.~Richter-Was, \href{https://www.sciencedirect.com/science/article/pii/037026939390062M?via\%3Dihub} {Phys. Lett. B {\bf 303}, 163 (1993)}.

% %\cite{Schmitz:1993qyg}
% \bibitem{Schmitz:1993qyg}
% W.~Schmitz, H.~H\"ubel, C.~X.~Yang, G.~Baldsiefen, U.~Birkental, G.~Fr\"ohlingsdorf, D.~Mehta, R.~M\"u\ensuremath{\beta}eler, M.~Neffgen and P.~Willsau, \textit{et al.}
% %``Transition quadrupole moments of a large-deformation intruder band in 163 Lu,''
% Phys. Lett. B \textbf{303}, 230-235 (1993)
% doi:10.1016/0370-2693(93)91425-M
% %41 citations counted in INSPIRE as of 29 May 2025

\bibitem{pkpi}
M.~Ablikim {\it et al.} (BESIII Collaboration),
\href{https://journals.aps.org/prl/abstract/10.1103/PhysRevLett.117.232002}{Phys. Rev. Lett. \textbf{120}, 029903 (2018)}.

% %\cite{BESIII:2016ozn}
% \bibitem{BESIII:2016ozn}
% M.~Ablikim \textit{et al.} [BESIII],
% %``Measurement of Singly Cabibbo Suppressed Decays $\Lambda_c^{+}\to p\pi^{+}\pi^{-}$ and $\Lambda_c^{+}\to pK^{+}K^{-}$,''
% Phys. Rev. Lett. \textbf{117}, no.23, 232002 (2016)
% doi:10.1103/PhysRevLett.117.232002
% [arXiv:1608.00407 [hep-ex]].
% %50 citations counted in INSPIRE as of 29 May 2025


\bibitem{golden}
M.~Ablikim {\it et al.} (BESIII Collaboration),
\href{https://journals.aps.org/prl/abstract/10.1103/PhysRevLett.116.052001}{Phys. Rev. Lett. \textbf{116}, 052001 (2016)}.


% %\cite{BESIII:2015bjk}
% \bibitem{BESIII:2015bjk}
% M.~Ablikim \textit{et al.} [BESIII],
% %``Measurements of absolute hadronic branching fractions of $\Lambda_{c}^{+}$ baryon,''
% Phys. Rev. Lett. \textbf{116}, no.5, 052001 (2016)
% doi:10.1103/PhysRevLett.116.052001
% [arXiv:1511.08380 [hep-ex]].
% %200 citations counted in INSPIRE as of 17 Jun 2025





\bibitem{fjh}
M.~Ablikim {\it et al.} (BESIII Collaboration),
\href{https://journals.aps.org/prd/abstract/10.1103/PhysRevD.109.L071104}{Phys. Rev. D {\bf 109}, L071104 (2024)}.

% %\cite{BESIII:2023eie}
% \bibitem{BESIII:2023eie}
% M.~Ablikim \textit{et al.} [BESIII],
% %``Measurements of Born cross sections for e+e-\textrightarrow{}\ensuremath{\Lambda}c+\ensuremath{\Lambda}\textasciimacron{}c(2595)-+c.c. and e+e-\textrightarrow{}\ensuremath{\Lambda}c+\ensuremath{\Lambda}\textasciimacron{}c(2625)-+c.c. at s=4918.0 and 4950.9~MeV,''
% Phys. Rev. D \textbf{109}, no.7, L071104 (2024)
% doi:10.1103/PhysRevD.109.L071104
% [arXiv:2312.08414 [hep-ex]].
% %3 citations counted in INSPIRE as of 29 May 2025

\bibitem{fvp-1}
F.~Jegerlehner, \href{https://link.springer.com/article/10.1007/BF01552495}{Z. Phys. C, \textbf{32}, 195 (1986)}.

% %\cite{Jegerlehner:1985gq}
% \bibitem{Jegerlehner:1985gq}
% F.~Jegerlehner,
% %``Hadronic Contributions to Electroweak Parameter Shifts: A Detailed Analysis,''
% Z. Phys. C \textbf{32}, 195 (1986)
% doi:10.1007/BF01552495
% %198 citations counted in INSPIRE as of 29 May 2025

\bibitem{fvp-2}
S.~Actis {\it et al.} (Working Group on Radiative Corrections, Monte Carlo Generators for Low Energies Collaboration), \href{https://link.springer.com/article/10.1140/epjc/s10052-010-1251-4}{Eur. Phys. J. C  \textbf{66}, 585 (2010)}.

% %\cite{WorkingGrouponRadiativeCorrections:2010bjp}
% \bibitem{WorkingGrouponRadiativeCorrections:2010bjp}
% S.~Actis \textit{et al.} [Working Group on Radiative Corrections and Monte Carlo Generators for Low Energies],
% %``Quest for precision in hadronic cross sections at low energy: Monte Carlo tools vs. experimental data,''
% Eur. Phys. J. C \textbf{66}, 585-686 (2010)
% doi:10.1140/epjc/s10052-010-1251-4
% [arXiv:0912.0749 [hep-ph]].
% %417 citations counted in INSPIRE as of 29 May 2025


\bibitem{fisr}
S.~Jadach, B.~F.~L.~Ward, and Z.~Was, \href{https://journals.aps.org/prd/abstract/10.1103/PhysRevD.63.113009}{Phys. Rev. D, \textbf{63}, 113009 (2001)}.


% %\cite{Jadach:2000ir}
% \bibitem{Jadach:2000ir}
% S.~Jadach, B.~F.~L.~Ward and Z.~Was,
% %``Coherent exclusive exponentiation for precision Monte Carlo calculations,''
% Phys. Rev. D \textbf{63}, 113009 (2001)
% doi:10.1103/PhysRevD.63.113009
% [arXiv:hep-ph/0006359 [hep-ph]].
% %879 citations counted in INSPIRE as of 29 May 2025


\bibitem{shapeLambadc}
M.~Ablikim {\it et al.} (BESIII Collaboration),
\href{https://journals.aps.org/prl/abstract/10.1103/PhysRevLett.131.191901}{Phys. Rev. Lett, \textbf{131}, 191901 (2023)}.

% %\cite{BESIII:2023rwv}
% \bibitem{BESIII:2023rwv}
% M.~Ablikim \textit{et al.} [BESIII],
% %``Measurement of Energy-Dependent Pair-Production Cross Section and Electromagnetic Form Factors of a Charmed Baryon,''
% Phys. Rev. Lett. \textbf{131}, no.19, 191901 (2023)
% doi:10.1103/PhysRevLett.131.191901
% [arXiv:2307.07316 [hep-ex]].
% %36 citations counted in INSPIRE as of 29 May 2025



\bibitem{else}
M.~Ablikim {\it et al.} (BESIII Collaboration),
\href{https://journals.aps.org/prl/abstract/10.1103/PhysRevLett.131.191901}{Phys. Rev. Lett, \textbf{126}, 102001 (2021)}.
% %\cite{BESIII:2020qkh}
% \bibitem{BESIII:2020qkh}
% M.~Ablikim \textit{et al.} [BESIII],
% %``Observation of a Near-Threshold Structure in the $K^+$ Recoil-Mass Spectra in $e^+e^- \rightarrow K^+(D_s^-D^{*0}+D_s^{*-}D^0$),''
% Phys. Rev. Lett. \textbf{126}, no.10, 102001 (2021)
% doi:10.1103/PhysRevLett.126.102001
% [arXiv:2011.07855 [hep-ex]].
% %255 citations counted in INSPIRE as of 17 Jun 2025


\bibitem{baye}
Y.~S.~Zhu,
\href{https://iopscience.iop.org/article/10.1088/1674-1137/32/5/007}{Chin. Phys. C \textbf{32}, 363 (2008)}.

% %\cite{Zhu:2008ca}
% \bibitem{Zhu:2008ca}
% Y.~S.~Zhu,
% %``Bayesian credible interval construction for Poisson statistics,''
% Chin. Phys. C \textbf{32}, 363 (2008)
% doi:10.1088/1674-1137/32/5/007
% [arXiv:0812.2705 [physics.data-an]].
% %21 citations counted in INSPIRE as of 29 May 2025


\bibitem{pid_sys}
M.~Ablikim {\it et al.} (BESIII Collaboration),
\href{https://journals.aps.org/prl/abstract/10.1103/PhysRevLett.112.251801}{Phys. Rev. Lett. \textbf{112}, 251801 (2014)}.

% %\cite{BESIII:2014bgm}
% \bibitem{BESIII:2014bgm}
% M.~Ablikim \textit{et al.} [BESIII],
% %``Observation of $\eta^{\prime}\to\pi^{+}\pi^{-}\pi^{+}\pi^{-}$ and $\eta^{\prime}\to\pi^{+}\pi^{-}\pi^{0}\pi^{0}$,''
% Phys. Rev. Lett. \textbf{112}, 251801 (2014)
% doi:10.1103/PhysRevLett.112.251801
% [arXiv:1404.0096 [hep-ex]].
% %35 citations counted in INSPIRE as of 29 May 2025






\bibitem{Barlow}
R.~Barlow, \href{https://arxiv.org/abs/hep-ex/0207026}{Systematic errors: Facts and fictions,  
[arXiv:hep-ex/0207026 [hep-ex]]}.
%217 citations counted in INSPIRE as of 03 Jun 2025






\bibitem{likehood}
X.~X.~Liu, X.~R.~L\"u, and Y.~S.~Zhu,
\href{https://iopscience.iop.org/article/10.1088/1674-1137/39/10/103001}{Chin. Phys. C \textbf{39}, 103001 (2015)}.

% %\cite{Liu:2015uha}
% \bibitem{Liu:2015uha}
% X.~X.~Liu, X.~R.~L\"u and Y.~S.~Zhu,
% %``Combined estimation for multi-measurements of branching ratio,''
% Chin. Phys. C \textbf{39}, no.10, 103001 (2015)
% doi:10.1088/1674-1137/39/10/103001
% [arXiv:1505.01278 [physics.data-an]].
% %25 citations counted in INSPIRE as of 29 May 2025




\bibitem{BESIII:2020nme}
M.~Ablikim {\it et al.} (BESIII Collaboration),
\href{https://iopscience.iop.org/article/10.1088/1674-1137/44/4/040001}{Chin. Phys. C \textbf{44}, 040001 (2020)}.

% %\cite{BESIII:2020nme}
% \bibitem{BESIII:2020nme}
% M.~Ablikim \textit{et al.} [BESIII],
% %``Future Physics Programme of BESIII,''
% Chin. Phys. C \textbf{44} (2020) no.4, 040001
% doi:10.1088/1674-1137/44/4/040001
% [arXiv:1912.05983 [hep-ex]].
% %678 citations counted in INSPIRE as of 25 Aug 2025




\bibitem{stcf}
M.~Achasov {\it et al.}, \href{https://link.springer.com/article/10.1007/s11467-023-1333-z} {Front. Phys.  \textbf{19},  14701 (2024)}.

% %\cite{Achasov:2023gey}
% \bibitem{Achasov:2023gey}
% M.~Achasov, X.~C.~Ai, R.~Aliberti, L.~P.~An, Q.~An, X.~Z.~Bai, Y.~Bai, O.~Bakina, A.~Barnyakov and V.~Blinov, \textit{et al.}
% %``STCF conceptual design report (Volume 1): Physics \& detector,''
% Front. Phys. (Beijing) \textbf{19}, no.1, 14701 (2024)
% doi:10.1007/s11467-023-1333-z
% [arXiv:2303.15790 [hep-ex]].
% %140 citations counted in INSPIRE as of 03 Jun 2025



\end{thebibliography}
\end{document}